\documentclass{LMCS}

\usepackage{color}
\usepackage{pstricks}
\usepackage{pst-node}
\usepackage{pst-tree}

\usepackage{amsmath}
\usepackage[amsfonts]{logic-thm}
\usepackage{enumerate,hyperref}

\theoremstyle{plain}\newtheorem{remark}{Remark}
\theoremstyle{plain}\newtheorem{lemma}[thm]{Lemma}
\renewcommand{\Aa}{\mathcal{A}}
\renewcommand{\Bb}{\mathcal{B}}
\renewcommand{\Pp}{\mathcal{P}}
\renewcommand{\Cc}{\mathcal{C}}

\newcommand{\vecs}{\vec s}
\newcommand{\vecsp}{\vec{s'}}
\newcommand{\vecspp}{\vec{s''}}
\newcommand{\vect}{\vec t}
\newcommand{\vecu}{\vec u}
\newcommand{\vecv}{\vec v}
\newcommand{\vecw}{\vec w}
\newcommand{\vAa}{\Aa_1\otimes\dots\otimes\Aa_n}
\newcommand{\vAaa}{\Aa'_1\otimes\dots\otimes\Aa'_n}
\newcommand{\vAap}{\Aa'_1\otimes\Aa''_1\dots\otimes\Aa'_n\otimes\Aa''_n}

\newcommand{\trans}[1]{\stackrel{#1}{\longrightarrow}}
\newcommand{\choice}{\zeta}
\newcommand{\cchoice}{\mathit{choice}}

\def\doi{4 (2:5) 2008}
\lmcsheading%
{\doi}
{1--14}
{}
{}
{Sep.~21, 2007}
{May~15, 2008}
{}   
\begin{document}

\title{A lower bound on web services composition}
\author[A.~Muscholl]{Anca Muscholl\rsuper a}
\address{{\lsuper a}LaBRI, Universit{\'e} Bordeaux\\
351, Cours de la Lib\'eration\\
F-33 405, Talence cedex, France}
\email{anca@labri.fr}

\author[I.~Walukiewicz]{Igor Walukiewicz\rsuper b}
\address{{\lsuper b}CNRS LaBRI, 351, Cours de la Lib\'eration,
F-33 405, Talence cedex, France}
\email{igw@labri.fr}

\thanks{{\lsuper{a,b}}Work supported by the projects
    ANR DOCFLOW (ANR-06-MDCA-005) and ANR DOTS (ANR-06-SETI-003).}



\begin{abstract}

A web service is modeled here as a finite state machine.  A composition
problem for web services is to decide if a given web service can be
constructed from a given set of web services; where the construction
is understood as a simulation of the specification by a fully asynchronous 
product of the given services. We show an EXPTIME-lower bound for this problem, 
thus matching the known upper bound.  Our result
also applies to richer models of web services, such as the Roman 
model.

\end{abstract}

\maketitle

\section{Introduction}

Inherently distributed applications such as web services~\cite{ACKM04}
increasingly get into the focus of automated verification
techniques. Often, some basic e-services are already implemented, but
no such simple service can answer to a more complex query. For
instance, a user interested in hiking Mt.~Everest will ask a travel
agency for information concerning weather forecast, group travels,
guides etc. The travel agency will contact different e-services,
asking for such information and making appropriate reservations, if
places are available. In general, single services such as weather
forecast or group reservations, are already available and it is
important to be able to reuse them without any change. The task of the
travel agency is to compose basic e-services in such a way that the
user's requirements are met (and eventually some constraints wrt.~the
called services, such as avoiding unreliable ones). Thus, one main
objective is to be able to check automatically that the composition of
basic e-services satisfies certain desirable properties or realizes
another complex e-service.

In this paper we study a problem that arises in the \emph{composition}
of e-services as considered in \cite{BCGLM03,BCGHM05,vldb05}. The
setting is the following: we get as input a specification (goal)
$\Bb$, together with $n$ available services $\Aa_1,\ldots,\Aa_n$. Then
we ask whether the composition of the services $\Aa_i$ can simulate
the behavior of the goal $\Bb$. This problem is known as
\emph{composition synthesis}. It amounts to synthesize a so-called
\emph{delegator}, that tells at any moment which service must perform
an action. In essence, a delegator implements a simulation relation of
the goal service $\Bb$ by the composition of the available services
$\Aa_i$.  In the most general setting, as considered for instance in
\cite{hbcs03,fbs04,dsv06}, services are modeled by communicating state
machines \cite{bz83}, that have access to some local data. In this
paper, we reconsider the simplified setting of the so-called Roman
model~\cite{BCGLM03} where services are finite state processes with no
access to data and no mutual synchronization. This restriction is
severe, however sufficient for our purposes, since our primary
motivation is to obtain a complexity lower bound for the composition
synthesis problem.

In this paper we study the complexity of the composition synthesis
problem in the very simple setting where the composition of the finite
state machines $\Aa_i$ is fully asynchronous (in particular there is
no communication). This case is interesting for two reasons. It is
known to be decidable in \EXPTIME~\cite{BCGLM03}, contrary to some
richer frameworks where it is undecidable~\cite{BCGHM05}. It is also
probably the simplest setting where the problem can be formulated,
thus the complexity of this variant gives a lower bound on the
complexity of any other variants of the synthesis problem.  A related
problem arises when instead of simulation one considers bisimulation.
This is sometimes called \emph{orchestration problem}, where the issue
is to find a communication architecture of the available services,
that is equivalent to the goal, modulo bisimulation. In our setting,
this problem amounts to checking if the asynchronous composition of
finite state machines is bisimilar to a given machine.

The main result of this paper is the \EXPTIME\ lower bound for the
composition synthesis problem. We also show that the same question can
be solved in polynomial time if we assume that the sets of actions of
the available machines are pairwise disjoint, i.e., each request can
be handled by precisely one service. Note that in the latter case, the
set of actions depends on the number of processes, whereas for the
first result we show that the case where the set of actions is fixed
is already \EXPTIME-hard. We also show that the orchestration
(bisimulation) problem
is \NLOGSPACE\ complete, independently of whether the sets of actions
of the components are disjoint or not\footnote{This problem is easier
  than checking bisimulation between a BPP and a finite state
  automaton, which is P-complete. The reason is that the finite-state
automaton is deterministic in our setting.}. This result, however, is less
interesting the context of service composition. The bisimulation
requirement means that that the client (goal automaton) should be
prepared to admit all possible interleavings in the composition, which
usually makes the specification too complex.

Similar kinds of questions were also considered by the verification
community. There is a large body of literature on the complexity of
bisimulation and simulation problems for different kinds of process
calculi (for a survey see \cite{Srba}). A result that is most closely
related to ours is the 
\EXPTIME\ completeness of simulation and bisimulation between non-flat
systems~\cite{LarSch00}. The main difference to our setting is that
there both a system and services are given as composition of finite state
machines using (binary) \emph{synchronization} on actions, i.e., an
action can synchronize two services. In a sense this paper shows that
the lower bound for the simulation holds even without any
synchronization.  

This paper is an extended version of the conference
publication~\cite{mw07}. In particular, the
characterization of the complexity of the bisimulation problem is new.

\section{Notations}
We denote throughout this paper tuples of states (i.e., global states
of a product automaton)  by bold characters
$\vec q, \vec s, \vec t, \dots$. Unless otherwise stated, the
components of  vector $\vec t$ are $t_1,\ldots,t_n$.

An \emph{asynchronous} product of $n$ deterministic automata 
\begin{equation*}
\Aa_i=\struct{Q_i,\S_i,q^0_i,\d_i :Q_i\times\S_i\to Q_i}  
\end{equation*}
 is a nondeterministic automaton:
\begin{equation*}
  \Aa_1\otimes\dots\otimes\Aa_n=\struct{Q,\S,\vec{q},\d: Q \times \S \to \Pp(Q)}
\end{equation*}
where: $Q= Q_1\times\cdots \times Q_n$; $\S=\bigcup_{i=1,\dots,n} \S_i$; 
$\vec{q}=(q^0_1,\dots,q^0_n)$; and $\d$ is defined by:
\begin{quote}
  $\vec t\in \d(\vec s,a)$ iff for some $i$, $t_i=\d_i(s_i,a)$ and for all 
  $j\not=i$ 
  we have $t_j=s_j$.
\end{quote}

Observe that the product automaton can be non deterministic because the alphabets $\S_i$ are not necessarily disjoint.

We define a \emph{simulation relation} on nondeterministic automata in
a standard way. Take two nondeterministic automata
$\Aa=\struct{Q_A,\S,q^0_A,\d_A:Q_A\times\S\to\Pp(Q_A)}$ and
$\Bb=\struct{Q_B,\S,q^0_B,\d_B:Q_B\times\S\to\Pp(Q_B)}$ over the same
alphabet. The simulation relation $\fleq\incl Q_A\times Q_B$ is the
biggest relation such that if $q_A\fleq q_B$ then for every $a\in \S$
and every $q'_A\in\d_A(q_A,a)$ there is $q'_B\in\d_B(q_B,a)$ such that
$q'_A\fleq q'_B$. We write $\Aa\fleq \Bb$ if $q^0_A\fleq q^0_B$.

\paragraph{Problem:} Given $n$ deterministic automata
$\Aa_1,\dots,\Aa_n$ and a deterministic automaton $\Bb$ decide if
$\Bb \fleq \Aa_1\otimes\dots\otimes\Aa_n$.

We will show that this problem is \EXPTIME-complete. It is clearly in
\EXPTIME\ as one can construct the product $\Aa_1\otimes\dots\otimes\Aa_n$
explicitly and calculate the biggest simulation relation with
$\Bb$. The rest of this paper will contain the proof of
\EXPTIME-hardness. We will start with the \PSPACE-hardness, as this will
allow us to introduce the method and some notation.

\section{A \PSPACE\  lower bound}

We will show \PSPACE-hardness of the problem by reducing it
to the  existence of a looping computation of a linearly space bounded
deterministic Turing machine. The presented proof of the \PSPACE\ bound 
has the advantage to generalize to the encoding of
alternating machines that we will present in the following section. 

Fix a deterministic Turing machine $M$ working in space bounded by the
size of its input. We want to decide if on a given input the
computation of the machine loops. Thus we do not need any accepting
states in the machine and we can assume that there are no transitions
from rejecting states. We denote by $Q$ the states of $M$ and by $\G$
the tape alphabet of $M$. A \emph{configuration} of $M$ is a word over
$\G\cup( Q\times\G)$ with exactly one occurrence of a letter from
$Q\times \G$. A configuration is of size  $n$ if it is a word of
length $n$. Transitions of $M$ will be denoted as $qa \trans{} q'bd$,
where $q,q'$ are the old/new state, $a,b$ the old/new tape symbol and
$d \in \{l,r\}$ the left/right head move (w.l.o.g.~we assume that $M$
moves the head in each step). 

Suppose that the input is a word $w$ of size $n$. We will construct
automata $\Aa_1,\dots,\Aa_n$ and $\Bb$ such that 
$\Bb \fleq \Aa_1\otimes\dots\otimes\Aa_n$ 
iff the computation of $M$ on $w$ is infinite. 

We start with some auxiliary alphabets. For every $i=1,\dots,n$ let
\begin{equation*}
\G_i=\G\times\set{i}\quad \text{and}\quad \D_i=(Q\times\G_i)\cup (Q\times\G_i\times\set{l,r})\, .
\end{equation*}
We will write $a_i$ instead of $(a,i)$ for elements of $\G_i$. Let also
$\D=\bigcup_{i=1,\dots,n} \D_i$.

The automaton $\Aa_i=\struct{Q_i,\S_i, q^0_i,\trans{}}$ is defined as follows:
\begin{enumerate}[$\bullet$]
\item The set of states is $Q_i=\G\cup (Q\times\G)\cup\set{\top}$, 
and the alphabet of the automaton is $\S_i=\D$.

\item We have transitions: 
  \begin{enumerate}[$-$]
  
  \item  $a\trans{qa_i}qa$, for all $a\in \G$ and $q\in Q$,

  \item $qa\trans{q'b_id} b$, for $qa\to q'bd$ the 
    transition of $M$ on $qa$ (there is at most one).

  \item From $a$, transitions on letters in $\D_i\setminus\set{qa_i :
      q\in Q}$ go to $\top$. Similarly, from $qa$ transitions on
    $\D_i\setminus \set{qb_id}$ go to $\top$ if there
    is a transition of $M$ on $qa$; if not, then $qa$ has
    no outgoing transitions. From $\top$ there are self-loops on
    all letters from $\D$.

  \end{enumerate}
\item For $i=2,\dots,n$ the initial state of $\Aa_i$ is $w_i$, the
  $i$-th letter of $w$; for $\Aa_1$ the initial state is $q^0w_1$,
  i.e., the initial state of $M$ and the first letter of $w$.
\end{enumerate}
Figure \ref{f:psp} shows a part of $\Aa_i$:
\vspace{1ex}

\begin{figure}[hbt]  
\centerline{\psset{nodesep=3pt,xunit=5cm,yunit=3cm,arrows=->}
\begin{pspicture}(2,1) 
  \cnodeput(0,0){t}{$\top$}
  \cnodeput(1,0){q}{$qa$}
  \cnodeput(1,1){a}{$a$}
  \cnodeput(2,1){b}{$b$}
  \ncline{a}{q}\naput{$qa_i$}
  \ncline{a}{t}\nbput[nrot=:D]{$\D_i\setminus\set{qa_i : q\in Q}$}
  \ncline{q}{t}\naput{$\D_i\setminus \set{q'b_id}$}
  \ncline{q}{b}\nbput{$q'b_id$}
  \nccircle[angleA=90,arrows=<-]{t}{.6cm}\naput{$\D$}
\end{pspicture}}
\vspace{2ex}
\caption{Part of $\Aa_i$}
         \label{f:psp}    
\end{figure}

\begin{figure}[htb]
\centerline{\psset{nodesep=3pt,xunit=5cm,yunit=3cm,arrows=->}
\begin{pspicture}(2,1)
\cnodeput(0,0.5){s}{$s$}
\cnodeput(2,0.5){e}{$e$}
\cnodeput(1,0){l}{$q'b_{i+1}l$}
\cnodeput(1,1){r}{$q'b_ir$}
\ncline{s}{l}\nbput{$q'b_{i+1}l$}
\ncline{s}{r}\naput{$q'b_ir$}
\ncline{l}{e}\nbput{$q'c_i$}
\ncline{r}{e}\naput{$q'c_{i+1}$}
\end{pspicture} 
} 
\vspace{2ex}

\caption{Automaton $K$}\label{f:K}
\end{figure}
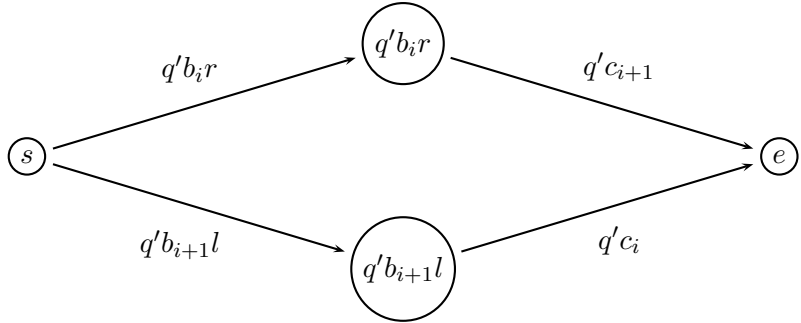

The idea is classical: automaton $\Aa_i$ controls the $i$-th tape
symbol, whereas automaton $\Bb$ defined below is in charge of the control part of
$M$. The challenge is to do this without using any synchronization
between adjacent automata $\Aa_i,\Aa_{i+1}$.
Next, we introduce an automaton $K$ that will be used to define
$\Bb$ (see also Figure \ref{f:K}). The set of states of $K$ is
$Q_K=\set{s,e}\cup (Q\times \bigcup 
\G_i\times\set{l,r})$; the initial state is $s$ and the final one $e$;
the alphabet is $\D$; the transitions are defined by:
\begin{enumerate}[$\bullet$]

\item $s\trans{q'b_ir} q'b_ir$ for $i=1,\dots,n-1$, whenever we have a
  transition $qa\to q'br$ in $M$ for some state $q$ and some letter $a$;
\item $s\trans{q'b_{i+1}l} q'b_{i+1}l$ for $i=1,\dots,n-1$, whenever we
  have a transition $qa\to qbl$ in $M$ for some state $q$ and some
  letter $a$; 

\item $q'b_ir \trans{q'c_{i+1}} e$ and $q'b_{i+1}l \trans{q'c_i} e$
  for all $c\in \G$.
\end{enumerate}
Figure \ref{f:K} presents a schema of the
automaton $K$.  We define $\Bb$ as the deterministic automaton
recognizing $(L(K))^*$, that is obtained by gluing together the states
$s$ and $e$.

\vspace{3ex}

\begin{remark}
  All $\Aa_i$ and $\Bb$ are deterministic automata of size polynomial in $n$. 
The input alphabets of the  $\Aa_i$ are almost pairwise disjoint: the only 
states with common labels on outgoing transitions are the $\top$ states.
\end{remark}

\begin{defi}
  We say that a configuration $C$ of size $n$ of $M$ \emph{corresponds} to a
  global state $\vecs$ of $\vAa$ iff $s_i=C(i)$ for $i=1,\dots,n$; in
  other words, if the state of $\Aa_i$ is the same as the $i$-th letter of
  $C$.
\end{defi}

\begin{defi}
  We say that a global state $\vecs$ of $\vAa$ is \emph{proper} when
  there is no $\top$-state in $\vecs$. 
\end{defi}

\begin{lemma}
  If $\vecs$ is a proper state, then for every letter $a\in\D$ the
  automaton $\vAa$ has in state $\vecs$ at most one outgoing
  $a$-transition.  Once the automaton enters a state that is not
  proper, it stays in non proper states.
\end{lemma}
 It is easy to see that from a non proper state, $\vAa$ can simulate 
any state of $\Bb$. The reason is that from $\top$,
 any move on letters from $\D$ is possible.

\begin{lemma} \label{l:step}
Suppose that $\vAa$ is in a state $\vecs$ that corresponds to a
  configuration $C$ of $M$.
  \begin{enumerate}[$\bullet$]
  \item If $C$ is a configuration with no successor,
    then there is a word $v\in L(K)$ that cannot be simulated by
    $\vAa$ from $\vecs$.

  \item Otherwise, the successor configuration $C\vdash C'$ exists,
    and there is a unique word $v\in L(K)$ such that
    $\vecs\trans{v}\vect$ and $\vect$ is proper. Moreover $\vect$
    corresponds to $C'$. All other words from $L(K)$ lead from $\vecs$
    to non
    proper states of $\vAa$. 
  \end{enumerate}
\end{lemma}

\begin{proof}

  For the first claim, assume that $\vecs$ corresponds to a configuration, 
thus there is exactly one $i$
  such that $\Aa_i$ is in a state from $Q\times\G$. The other
  automata are in states from $\G$. 
  
  If $C$ is terminal then $\Aa_i$ is in a state $qa$ which has
  no  outgoing transition. This means that this state can
  simulate no  move on letters $q'b_ir$, for  $q'\in Q$ and $b_i \in \G_i$ (and
  such a move exists in $K$, as the machine $M$ must have a move to the
  right if it is nontrivial). All other automata are also not capable
  to simulate $q'b_ir$ as they can do only moves on letters $\D_j$ for
  $j\not=i$.

  Now suppose that $C\vdash C'$. To avoid special, but simple, cases
  suppose that the position $i$ of the state is neither the first nor
  the last. Let $s_i=qa$ and suppose also that $qa\to q'br$ is the
  move of $M$ on $qa$. The case when the move is to the left is
  similar.

  The only possible move of $K$ from $s$ which will put $\vAa$
  into a proper state is $q'b_ir$. This makes $\Aa_i$ to change the
  state to $b$ and it makes $K$ to change the state to $q'b_ir$. From
  this latter state the only possible move of $K$ is on letters
  $q'c'_{i+1}$ for arbitrary $c'\in \G$. Suppose that $\Aa_{i+1}$ is in the
  state $c=s_{i+1} \in \G$, then all moves of $K$ on $q'c'_{i+1}$ with $c'\not=c$
  can be matched with a move to $\top$ of $\Aa_{i+1}$. On $q'c_{i+1}$
 the automaton $\Aa_{i+1}$ goes to $q'c$ and automaton $K$ goes to
  $e$. This way the state in the configuration is changed and
  transmitted to the right. We have that the new state of $\vAa$
  corresponds to the configuration $C'$.

\end{proof}

\begin{lemma}
  We have $\Bb \fleq \vAa$ iff the computation of $M$ on $w$ is
  infinite. 
\end{lemma}

\begin{proof}
  Recall that $\Bb$ is a deterministic automaton recognizing
  $(L(K))^*$, and has initial state $s$.  
The initial state of $\vAa$ corresponds to the initial
  configuration $C_0$ of $M$ on $w$. We show now for every state 
$\vect$ corresponding to a configuration $C$ of $M$:  $s \fleq \vect$ iff the 
computation of $M$ starting in $C$ is infinite.

From a configuration $C$, the machine $M$ has only one computation: either
infinite, or a finite one that is blocking. Suppose that the computation
from $C$ has at least one step and let $C_1$ be the successor
configuration. By Lemma~\ref{l:step} from state $s$ there is exactly
one word $v_1\in L(K)$ such that $\vect \act{v_1} \vect_1$ in $\vAa$,
and $\vect_1$ is proper.  Moreover $\vect_1$
corresponds to $C_1$. On all other words from $L(K)$, the product
$\vAa$ reaches non proper states and from there it can simulate any
future behaviour of $\Bb$.  If $C_1$ has no successor configuration
then, again by Lemma~\ref{l:step}, there is a word in $L(K)$ that
cannot be simulated by $\vAa$ from $\vect_1$. If $C_1$ has a successor
then we repeat the whole argument. Thus the behaviour of $\Bb$ from $s$
can be simulated by $\vAa$ from the state corresponding to $C$
iff the machine $M$ has an infinite computation starting from $C$.

\end{proof}

\medskip

One can note that the construction presented in this section uses
actions that are common to several processes in a quite limited way:
the only states that have common outgoing labels are the $\top$ states
from which all behaviours are possible. This observation motivates the
question about the complexity of the problem when the
automata $\Aa_1,\dots,\Aa_n$ have pairwise disjoint alphabets. With
this restriction, the simulation problem can be solved efficiently:

\begin{thm}
The following question can be solved in polynomial time:

\emph{Input:} $n$ deterministic automata
$\Aa_1,\dots,\Aa_n$ over pairwise disjoint input alphabets, 
and a deterministic automaton $\Bb$.

\emph{Output:} decide if
$\Bb \fleq \Aa_1\otimes\dots\otimes\Aa_n$.
\end{thm}

\begin{proof}
  Let $\Cc_i$ be a automaton with a single state $\top$, and with
  self-loops on every letter from the alphabet $\S_i$ of $\Aa_i$. We
  write $\Aa^{(i)}$ for the asynchronous product of all $\Cc_j$, $j
  \not= i$, and of $\Aa_i$. Similarly, $\vect^{(i)}$ will denote $\vect$
with all components but $i$ replaced by $\top$.  Suppose now that $p$
is a state of $\Bb$, and $\vect$ a state of $\vAa$.  We write $p
\fleq_i \vect$ if $p$ is simulated by $\vect^{(i)}$ in $\Aa^{(i)}$.
 Notice that since $\Bb$ and $\Aa_i$ are both deterministic,
we can decide if $p \not\fleq_i \vect$ in logarithmic space (hence in
polynomial time), by guessing simultaneously a path in $\Bb$ and one
in $\Aa_i$.

  We show now that $p \fleq \vect$ in $\vAa$ iff $p \fleq_i \vect$ for
  all $i$.

  If $p \fleq \vect$, then all the more $p \fleq
  \vect^{(i)}$, since $\Cc_j$ can simulate $\Aa_j$ for all $j=1,\dots,n$.
  Conversely, assume that $p \fleq_i \vect$
  for all $i$, but  $p \not\fleq \vect$. This means that
  there exist computations $p \trans{a_1\ldots a_k} p'$ in $\Bb$,
  $\vect \trans{a_1\ldots a_k} \vecu$ in $\vAa$ and a letter $a
  \in\S_i$ for some $i$, such that $p'$ has an outgoing
  $a$-transition, but $\vecu_i$ does not (in $\Aa_i$). Clearly, we
  also have a computation $\vect^{(i)} \trans{a_1\ldots a_k}
  \vecu^{(i)}$ in $\Aa^{(i)}$. Since $\vecu_i$ has no outgoing
  $a$-transition, so neither does $\vecu^{(i)}$, which contradicts $p
  \fleq_i \vect$.
\end{proof}

\section{The complexity of simulation}

This time we take an alternating Turing machine $M$ working in space
bounded by the size of the input. We want to decide if $M$ has an
infinite computation.  This means that the machine can make choices of
existential transitions in such a way that no matter what are the
choices of universal transitions the machine can always
continue. Clearly, one can reduce the word problem to this problem,
hence it is \EXPTIME-hard (see \cite{cks81}; for more details on
complexity see any standard textbook).

We will assume that $M$ has always a choice between two transitions,
i.e., for each non blocking state/symbol pair $qa$ there will be precisely two 
distinct tuples
$q'b'd'$, $q''b''d''$ such that $qa\to q'b'd'$ and $qa\to
q''b''d''$. If $q$ is existential then it is up to the machine to choose
a move; if $q$ is universal then the choice is made from 
outside. To simplify the presentation we will assume that $d'=d''$,
i.e., both moves go in the same direction.  Every machine can be
transformed to an equivalent one with this property. We will also
assume that the transitions are ordered in some way, so we will be able
to say that $qa\to q'b'd$ is the first transition and $qa\to q''b''d$
is the second one.

Take the input word is $w$ of size $n$. We will construct
automata $\Aa'_1,\Aa''_1,\dots,\Aa'_n,\Aa''_n$ and $\Bb$ such that $\Bb$
is simulated by $\vAap$ iff there is an infinite alternating
computation of $M$ on $w$. The main idea is that automata $\Aa'_i$ and $\Aa''_i$ 
control the $i$-th tape symbol, as in the previous section, and each one is in charge 
of one of the two possible transitions (if any) when the input head is
at position $i$ in an existential state (universal moves are simpler).

We will modify a little the alphabets that we use. Let
\begin{align*}
  \D'_i=&(Q\times\G_i)\cup (Q\times\G_i\times\set{l,r}\times\set{1})\\
  \D''_i=&(Q\times\G_i)\cup (Q\times\G_i\times\set{l,r}\times\set{2})
\end{align*}
We then put $\D_i=\D'_i\cup\D''_i$, $\D=\bigcup_i \D_i$, 
$\D' = \bigcup_i \D'_i$ and $\D'' = \bigcup_i \D''_i$.

The automaton $\Aa'_i$ is defined as follows:
\begin{enumerate}[$\bullet$]
\item The set of states is $Q'_i=\set{\top}\cup\G\cup (Q\times\G)\cup
  (Q\times\G\times\set{l,r})$, the alphabet of the automaton is
  $\S'_i=\D\cup\set{\choice}$; where $\choice$ is a new letter common
  to all  automata.

\item We have the following transitions: 
  \begin{enumerate}[$-$]
  \item $a\trans{qa_i} qa$ for all $a\in \G$ and $q\in Q$,

  \item $qa\trans{q'b'_id1} b'$ and $qa\trans{q''b''_id1} b''$ if $q$ is
    an universal state and $qa\to q'b'd$, $qa\to q''b''d$ are the two
    transitions from $qa$. We have also transitions to $\top$ on all
    the letters from $\D'_i\setminus \set{q'b'_id1,q''b''_id1}$.

  \item $qa\trans{\choice} q'b'd\trans{q'b'_id1}b'$ and
    $qa\trans{q''b''_id1} b''$ if $q$ is an existential state and
    $qa\to q'b'd$, $qa\to q''b''d$ are the first and the second
    transitions from $qa$, respectively. We have also transitions to
    $\top$ on all the letters from $\D'_i\setminus \set{q''b''_id1}$. 
     From $q'b'd$ all transitions
    on $\D'_i \setminus \set{q'b'_id1}$ go to $\top$.

  \item From $a$, transitions on letters in $\D'_i \setminus \set{qa_i
      : q\in Q}$ go to $\top$.  If $qa$ is terminal then there are no
    outgoing transitions from $qa$.  From $\top$ there are self-loops
    on all letters from $\D^c := \D\cup\set{\choice}$.
  \end{enumerate}

\item The initial state of $\Aa'_i$ is $w_i$, the $i$-th letter
  of $w$ except for $\Aa_1$ whose initial state is $q^0w_1$, the
  initial state of $M$ and the first letter of $w$.
\end{enumerate}
Figure~\ref{fig:Aprim} below presents parts of $\Aa'_i$ corresponding to
universal and existential states.
\vspace{2ex}

\begin{figure}[htb]
  \centering
{
\psset{nodesep=3pt,xunit=5.6cm,yunit=3cm,arrows=->}
\begin{pspicture}(1.9,1)
  \cnodeput(0,0){t}{$\top$}
  \cnodeput(1,0){q}{$qa$}
  \cnodeput(1,1){a}{$a$}
  \cnodeput(1.9,1){bp}{$b'$}
  \cnodeput(1.4,1){bpp}{$b''$}
  \ncline{a}{q}\naput{$qa_i$}
  \ncline[arrows=<-]{t}{a}\naput[nrot=:U]{$\D'_i\setminus \set{qa_i : q\in Q}$}
  \ncline{q}{t}\naput{$\D'_i\setminus \set{q'b'_id1,q''b''_id1}$}
\psset{arcangle=-20}
  \ncarc{q}{bp}\nbput[nrot=:U]{$q'b'_id1$}
  \ncarc{q}{bpp}\nbput[nrot=:U]{$q''b''_id1$}
  \nccircle[angleA=90,arrows=<-]{t}{.5cm}\naput{$\D^c$}
\end{pspicture}
}\vspace{6ex}

{\psset{nodesep=3pt,xunit=5.6cm,yunit=3cm,arrows=->}
\begin{pspicture}(1.9,1)
  \cnodeput(0,0){t}{$\top$}
  \cnodeput(1,0){q}{$qa$}
  \cnodeput(1.5,.4){c}{$q'b'd$}
  \cnodeput(1,1){a}{$a$}
  \cnodeput(1.9,1){bp}{$b'$}
  \cnodeput(1.4,1){bpp}{$b''$}
  \cnodeput(2,0){tp}{$\top$}
  \ncline{a}{q}\naput{$qa_i$}
  \ncline[arrows=<-]{t}{a}\naput[nrot=:U]{$\D'_i\setminus \set{qa_i : q\in Q}$}
  \ncline{q}{t}\naput{$\D'_i\setminus \set{q''b''_id1}$}
  \psset{arcangle=-20}
  \ncarc{q}{c}\nbput[nrot=:U]{$\choice$}
  \ncarc{c}{bp}\nbput[nrot=:U]{$q'b'_id1$}
  \ncarc{q}{bpp}\nbput[nrot=:U]{$q''b''_id1$}
  \nccircle[angleA=90,arrows=<-]{t}{.5cm}\naput{$\D^c$}
  \ncline{c}{tp}\nbput[nrot=:U]{$\D'_i\setminus \set{q'b'_id1}$}
\end{pspicture}
}
\vspace{3ex}

\caption{Parts of the automaton $\Aa'_i$ corresponding to universal
  and existential states $q$, respectively. The alphabet $\D^c$ is
  $\D\cup\set{\choice}$.}
  \label{fig:Aprim}
\end{figure}
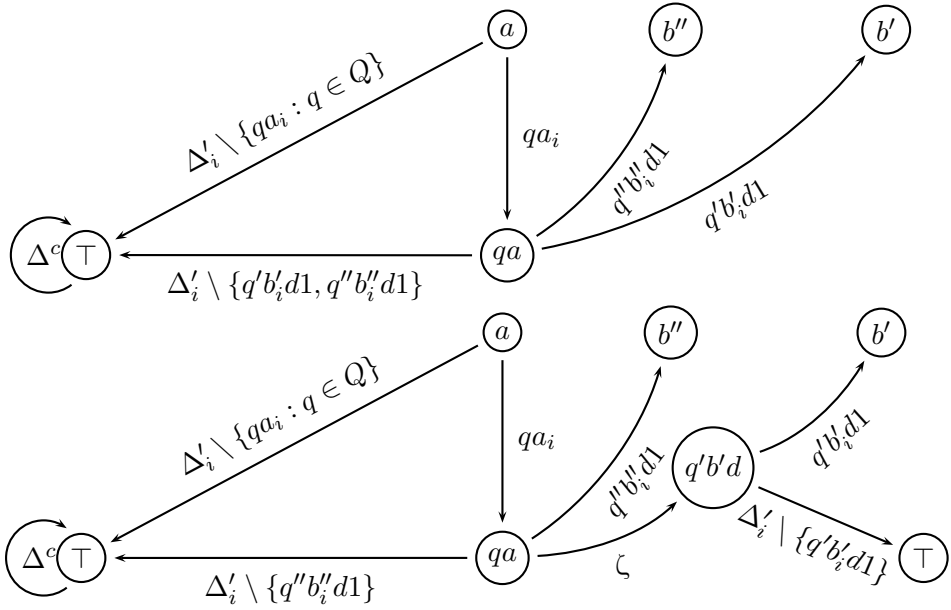

The automaton $\Aa''_i$ is the same as $\Aa'_i$ with the difference
that we replace every label $q''b''d1$ by $q'b'd2$,
every $q'b'd1$ by $q''b''d2$ (notice the change of primes and double
primes), every  $\D'_i$ by $\D''_i$ and $\D'$ by $\D''$. Moreover,
state labels $b'$ and $b''$ are exchanged, and state $q'b'd$ is
relabeled $q''b''d$.

\begin{figure}[htb]
  \centering%
{\psset{nodesep=3pt,xunit=5.4cm,yunit=3cm,arrows=->}
  \begin{pspicture}(2,2)
  \cnodeput(0,1){l}{$q'b_{i+1}l$} 
  \cnodeput(2,1){r}{$q'b_ir$} 
  \cnodeput(1,2){s}{$s$}  
  \cnodeput(1,1){c}{$\cchoice$}  
  \cnodeput(1,0){e}{$e$}  
  \ncline{s}{l}\nbput[nrot=:D]{$(q'b_{i+1}l1)(q'b_{i+1}l2)$}
  \ncline{s}{r}\naput[nrot=:U]{$(q'b_ir1)(q'b_ir2)$}
  \ncline{c}{l}\nbput[nrot=:D]{$(q'b_{i+1}l1)(q'b_{i+1}l2)$}
  \ncline{c}{r}\naput[nrot=:U]{$(q'b_ir1)(q'b_ir2)$}
  \ncline{s}{c}\nbput[nrot=:L]{$\choice$}
  \ncline{l}{e}\nbput[nrot=:U]{$(q'c_i)(q'c_i) $}
  \ncline{r}{e}\naput[nrot=:D]{$(q'c_{i+1})(q'c_{i+1})$}
  \end{pspicture}
}
  
  \caption{Automaton $K$}
  \label{fig:autK}
\end{figure}
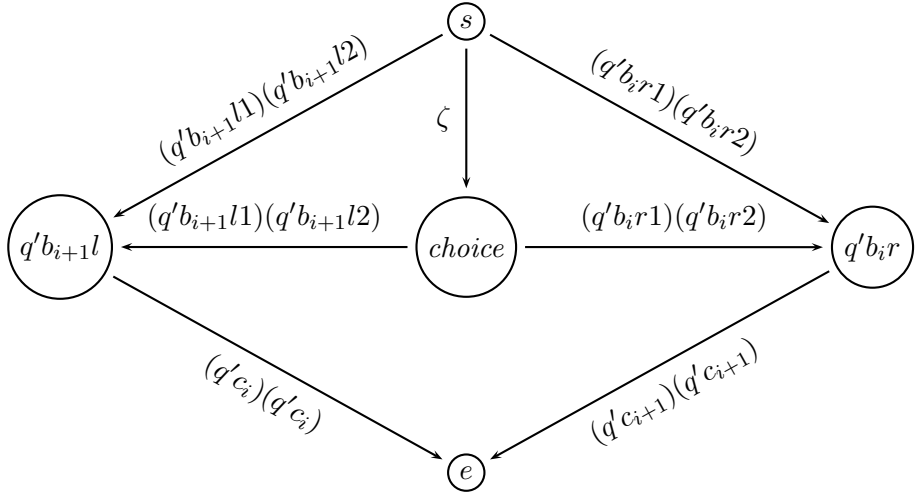

Next, we define a new automaton $K$ that will be used to define new
automaton $\Bb$. The states of $K$ are
\begin{equation*}
Q_K=\set{s,e,\cchoice}\cup (Q\times\bigcup_i\G_i\times\set{l,r})
\end{equation*}
plus some auxiliary states to implement transitions on two letters at
a time. We will write transitions with two letters on them for
readability. The initial state is $s$ and the final one is $e$. The
alphabet is $\S_K=\bigcup \S_i$. The transitions are defined by (cf.\
Figure~\ref{fig:autK}):
\begin{enumerate}[$\bullet$]

\item $s\trans{\choice}\cchoice$;

\item $s\trans{(q'b_ir1)(q'b_ir2)} q'b_ir$ whenever we have a
  transition $qa\to q'br$ in $M$ for some universal state $q$ and 
some letter $a$, and
  similarly from $\cchoice$ instead of $s$ when $q$ is existential;

\item $s\trans{(q'b_{i+1}l1)(q'b_{i+1}l2)} q'b_{i+1}l$ whenever we
  have a transition $qa\to q'bl$ in $M$ for some universal state $q$
  and some letter $a$, and similarly from $\cchoice$ instead of $s$
  when $q$ is existential;

\item $q'b_ir \trans{(q'c_{i+1})^2} e$ and $q'b_{i+1}l \trans{(q'c_i)^2} e$
  for all $c\in \G$. 
\end{enumerate}
We define $\Bb$ as the deterministic automaton recognizing
$(L(K))^*$ that is obtained by gluing together states $s$ and $e$.
\vspace{2ex}

\begin{remark}
  All $\Aa'_i$, $\Aa''_i$ and $\Bb$ are deterministic and of size
  polynomial in $n$.
\end{remark}

\begin{defi}\label{df:cor}
  A configuration $C$ of size $n$ \emph{corresponds} to a global state
    $\vec s$ of $\vAap$ if $s_{2i}=s_{2i-1}=C(i)$  for $i=1,\dots,n$;
    in other words, if the states of $\Aa'_i$ and $\Aa''_i$ are the
    same as the $i$-th letter of $C$.
\end{defi}

\begin{defi}
  We say that a global state $\vecs$ of $\vAap$ is \emph{proper}
  when $\top$ does not appear in $\vecs$.
\end{defi}

It is easy to see that from a non proper state, $\vAap$ can simulate 
any state of $\Bb$. The reason is that from $\top$,
 any move on letters from $\D^c$ is possible.

\begin{lemma}\label{lemma:mainexp}
  Suppose that $\vAap$ is in a state $\vec s$ corresponding to a
  configuration $C$ of $M$. If $C$ has no successor configuration
  then there is a word $v\in
  L(K)$ that cannot be simulated by  $\vAap$ from $\vecs$. Otherwise, $C$ has two
  successor configurations $C\vdash C'$ and $C\vdash C''$. We have two cases:
  \begin{enumerate}[$\bullet$]

  \item If $C$ is universal then there are two words $v'$ and $v''$ in
    $L(K)$: each leading from $\vecs$ to a unique state $\vect'$ and
    $\vect''$, respectively. These two states are proper and
    correspond to $C'$ and $C''$, respectively. On all other words
    from $L(K)$, non proper states can be reached from $\vecs$.

  \item If $C$ is existential, then on the letter $\choice$ exactly
    two states are reachable from $\vecs$, call them $\vecsp$ and
    $\vecspp$. There is a word $v'$ such that $\choice v' \in L(K)$
    and on $v'$ from $\vecsp$ a unique state is reachable. This state
    is proper and corresponds to $C'$. Similarly there is a word $v''$
    for $\vecspp$ and $C''$. On all words from $L(K)$ that are different from $\choice
    v'$ and $\choice v''$, non proper states can be reached from
    $\vecs$.
  \end{enumerate}
\end{lemma}

\begin{proof}
  As $\vecs$ corresponds to the configuration $C$, there is some $i$
  such that both automata $\Aa'_i$ and $\Aa''_i$ are in state $qa$,
  for some $q \in Q$ and $a \in \G$, and all other automata are in
  states from $\G$.

  If $C$ is a configuration without successor, then the state $qa$ in
  $\Aa'_i$ and $\Aa''_i$ does not have any outgoing transition. Thus
  these automata cannot simulate the $\choice$ transition of $K$ from
  $s$. No other automaton $\Aa'_j$, or $\Aa''_j$ can simulate the
  $\choice$ transition either, as they are all in states from $\G$.

  Suppose that $C$ is an universal configuration with two possible
  transitions to the right, $qa\to q'b'r$ and $qa\to q''b''r$. The
  case when the moves are to the left is similar. In $\Aa'_i$ from the
  state $qa$ we have a transition on $q'b'_ir1$ leading to $b'$ and on
  $q''b''_ir1$ leading to $b''$. Similarly for $\Aa''_i$, but on
  $q'b'_ir2$ and $q''b''_ir2$. These transitions can simulate both
  transitions $(q'b'_ir1)(q'b'_ir2)$ and $(q''b''_ir1)(q''b''_ir2)$
  that are possible from $s$ in $K$. (All other transitions from $s$ in
  $K$ lead from $\vecs$ to a non proper state of $\vAap$.) Let us focus
  only on the first case, when $(q'b'_ir1)(q'b'_ir2)$ is executed in $K$
  and the state $q'b'_ir$ is reached. From this state only 
  transitions $(q'c'_{i+1})^2$ are possible, for all $c'\in
  \G$. Suppose that $\Aa'_{i+1}$ and $\Aa''_{i+1}$ are in state
  $c\in\G$. Transition $(q'c_{i+1})^2$ of $K$ is simulated by moves
  to $q'c$ in both $\Aa'_{i+1}$ and $\Aa''_{i+1}$.  This way the new
  state is transferred to the right.  Transitions $(q'c'_{i+1})^2$
  where $c\not=c'$ are simulated in $\vAap$ by moves of $\Aa'_{i+1}$
  and $\Aa''_{i+1}$ to $\top$. 

  Suppose that $C$ is an existential configuration, with possible
  transitions $qa\to q'b'r$ and $qa\to q''b''r$. The case when moves
  are to the left is similar.  Consider first the transition of $K$
  from $s$ that corresponds to the letter $\choice$.  Both $\Aa'_i$
  and $\Aa''_i$ can simulate this transition: the first goes to state
  $q'b'r$, and the second goes to $q''b''r$. Assume that it is the
  transition of $\Aa'_i$ that is taken; the other case is
  symmetric. We get to the position when $K$ is in the state
  $\cchoice$, $\Aa'_i$ is in the state $q'b'r$ and $\Aa''_i$ in the
  state $qa$. From $\cchoice$, automaton $K$ can do
  $(q'b'_ir1)(q'b'_ir2)$ that can be simulated by the transitions of
  $\Aa'_i$ and $\Aa''_i$ (every other transition of $K$ can be
  simulated by a move of $\vAap$ to a non proper state).  Both
  automata reach the state $b'$. Automaton $K$ is now in state
  $q'b_ir$ from where it can do $(q'c_{i+1})^2$ for any $c\in \G$. The
  result of simulating these transitions while reaching a proper state
  is the transfer of the state to the right, in the same way as in the
  case of the universal move. Finally, it remains to see what happens
  if $K$ makes a move from $s$ that is different from $\choice$.  In
  this case, at least one of the automata $\Aa'_i$, $\Aa''_i$ can
  simulate the corresponding transition on $(p e_id1)$, $(p e_id2)$
  respectively, by going to state $\top$, since we suppose that in any
  configuration of $M$, the two outgoing transitions are
  distinct. Hence, a non proper state can be reached.
\end{proof}

\begin{thm} \label{th:main}
  The following problem is \EXPTIME-complete:

\emph{Input:} deterministic automata
$\Aa_1,\dots,\Aa_n$  and a deterministic automaton $\Bb$.

\emph{Output:} decide if
$\Bb \fleq \Aa_1\otimes\dots\otimes\Aa_n$.

\end{thm}
\begin{proof}
  The problem is clearly in \EXPTIME\ as the state space of $\vAap$
  can be constructed in \EXPTIME. For \EXPTIME\ hardness, we take an
  alternating machine $M$ as at the beginning of this section and use
  the construction presented above together with
  Lemma~\ref{lemma:mainexp}. Recall, that $\Bb$ is a deterministic
  automaton obtained from the automaton $K$ by gluing states $s$ and
  $e$ (cf.\ Figure~\ref{fig:autK}). We also have that the initial
  state of $\vAap$ corresponds to the initial configuration of $M$ (in
  a way required by Definition~\ref{df:cor}). We will show that for
  every state $\vect$ corresponding to a configuration $C$ of $M$: $s\fleq
  \vect$ iff $M$ has an infinite alternating computation from $C$.

  Consider a game of two players: Computer and Environment. Positions
  of the game are configurations of $M$. In existential configurations
  Computer chooses a successor configuration (with respect to the
  transition table of $M$). In universal configurations Environment
  makes a choice. Having an infinite alternating computation from $C$
  is equivalent to saying that in this game Computer has a strategy to
  avoid being blocked. At the same time, not having such a computation
  from $C$ is equivalent to saying Environment has a strategy to reach
  a configuration with no successors. As this is a reachability game,
  for each such $C$ there is a bound $d_C$ (distance) on the number of
  steps in which Environment can force Computer into a blocking
  configuration. This distance is $0$ if $C$ is blocking; it is one
  plus the maximum over distances for two successor configurations if
  $C$ is existential, and it is one plus the minimum over the
  distances of successor configurations if $C$ is universal. (Here we
  assume that the distance is $\infty$ if Environment cannot win from
  $C$.).

  Going back to the proof of the theorem, consider first the case when
  $M$ does not have an infinite alternating computation from $C$. Let
  $\vect$ be the state of $\vAap$ corresponding to $C$.  We show that
  $s\not\fleq \vect$ by induction on the distance $d_C$. There are
  three possible cases:
  \begin{enumerate}[$\bullet$]
  \item If $d_c=0$ then is no transition possible from $C$. In this case 
    Lemma~\ref{lemma:mainexp} gives us an execution of $\Bb$ from $s$ that
    cannot be simulated by $\vAap$ from $\vect$.
  \item If $C$ is universal, there is a successor $C_1$ such that
    $d_C>d_{C_1}$. We take the word $v\in L(K)$ given by
    Lemma~\ref{lemma:mainexp}. The only way to simulate this word from
    $\vect$ leads to the proper state $\vect_1$ corresponding to
    $C_1$. By induction hypothesis $s\not\fleq \vect_1$.
  \item If $C$ is existential, then for both successor configurations,
    $C'$ and $C''$, the distance is smaller. We make $\Bb$ execute
    $\choice$ and then, depending how it was matched by $\vAap$ , a
    word forcing the automaton to go to a proper state corresponding
    either to $C'$ or to $C''$. Using the induction hypothesis we get that
    the simulation is not possible from $s$ and the obtained states.
  \end{enumerate}

  The case when $M$ has an infinite alternating computation from $C$
  is very similar. In this case $d_C=\infty$. The means that if $C$ is
  an existential computation then one of the successor configurations
  has distance equal to $\infty$. By Lemma~\ref{lemma:mainexp} we can
  match $\choice$ so that we go to the state corresponding to that
  configuration. If $C$ is universal then both successor
  configurations have distance equal to $\infty$. Once again
  Lemma~\ref{lemma:mainexp}, tells us how to match every word from
  $L(K)$.
\end{proof}

We conclude the section by showing that Theorem \ref{th:main} still
holds under the assumption that the alphabet of the automata $\Aa_i$
and $\Bb$ is of
constant size.

\begin{thm} \label{th:mainc} Let $\S$ be a fixed alphabet of at
  least $2$ letters. The following problem is \EXPTIME-complete:

\emph{Input:} deterministic automata
$\Aa_1,\dots,\Aa_n$  and a deterministic automaton $\Bb$ over
the input alphabet $\S$.

\emph{Output:} decide if
$\Bb \fleq \Aa_1\otimes\dots\otimes\Aa_n$.

\end{thm}

\begin{proof}
  We reduce directly from Theorem \ref{th:main}. Suppose that
  the input alphabet of all automata $\Aa_i,\Bb$ is $\S\times
  \{1,\ldots,m\}$, for some $m$. Moreover,
  let $S$ be the set of states of $\Bb$ and let $Q = Q_1 \times \cdots
  \times Q_n$ be the set of global states of $\vAa$.

  In each automaton $\Aa_i$, $\Bb$ we replace every transition $s
  \trans{a_l} t$ by a sequence of transitions with labels from $\S
  \cup \{\#,\$\}$ as follows:
\[
s \trans{a} (stl0) \trans{\#} (stl1) \trans{\#} (stl2) \cdots
\trans{\#} (stll) \trans{\$} t
\]
The $(l+1)$ states $(stl0),\ldots, (stll)$ are new. Let $\Aa'_i,\Bb'$
be the automata obtained from  $\Aa_i$, $\Bb$, with state space $Q'$ and $S'$,
respectively.

Take $\fleq$, the largest simulation relation from $\Bb$
to $\vAa$. We show how to extend $\fleq$ to $\fleq'$
such that $\fleq'$ is a simulation relation from $\Bb'$
to $\vAaa$ (not necessarily the largest one). Let $\fleq'$ be the union of $\fleq$ with 
the set of all pairs $((stlk), \vecu')$, where 
$s,t \in S$, $\vecu'=(u'_1,\ldots,u'_n) \in Q'$, and
such that:

\begin{enumerate}[$\bullet$]
\item $s \trans{a_l} t$ and $\vecv \trans{a_l} \vecw$ for some $a\in \S$,
  $\vecv=(v_1,\dots,v_n)$ and $\vecw=(w_1,\dots,w_n)$ such that
  $s\fleq \vecv$, $t\fleq \vecw$,

\item there is some $i$ with $u'_i=(v_iw_ilk)$, and $u'_j=v_j=w_j$ for
  $j\not=i$.
\end{enumerate}
It is immediate to check that $\fleq'$ is a simulation relation.
First, (old) states from $S$ can only be simulated by (old) states
from $Q$.  Second, a new state $(stlj)$ of $\Bb$ can be simulated only
by states $\vecu' \in Q' \setminus Q$. It can be shown easily that the
largest simulation relation from $\Bb'$ to $\vAaa$ coincides with
$\fleq'$ (hence with $\fleq$) on the set $S \times Q$ of pairs of old
states. 

\end{proof}

\section{The complexity of bisimulation}

Till now we  wanted to decide if an asynchronous product of
deterministic automata $\vAa$ can simulate a
deterministic automaton $\Bb$. An evident question is to consider what
happens if we consider bisimulation instead  of simulation. To be
bisimilar to an asynchronous product, $\Bb$ must satisfy some 
structural constraints. In this section we prove the following theorem,
which shows that indeed, the bisimulation problem is easier. 

\begin{thm}
  The following question can be solved in logarithmic space:

  \emph{Input:} $n$ deterministic automata $\Aa_1,\dots,\Aa_n$ and a
  deterministic automaton $\Bb$.

  \emph{Output:} decide if $\Bb$ and $\vAa$ are bisimilar.
\end{thm}

The proof of the theorem will occupy the rest of the section. We fix
$\Bb$ and $\Aa_1,\dots,\Aa_n$. Without loss of generality we assume
that $\Bb$ is minimal with respect to bisimulation: no two different
states of $\Bb$ are bisimilar (if $\Bb$ is not minimal we can
minimize it on-the-fly in logarithmic space).  This
assumption also has a very pleasant consequence. If two states $s_1$
and $s_2$ of $\Bb$ are bisimilar to the same global state of $\Aa$,
then $s_1=s_2$.

As we aim to obtain a logarithmic space algorithm we cannot even allow
ourselves to explore the state space of $\vAa$ at random, as we cannot
store the tuples of states. This is why the following definition is
crucial for the construction.

\begin{defi}
  A sequence of transitions of $\vAa$ is
  \emph{banal} if it can be decomposed into a, possibly empty,
  sequence of transitions of $\Aa_1$, followed by one of $\Aa_2$,
  and so on, up to $\Aa_n$.
\end{defi}

Observe that thanks to the lack of synchronization every state of
$\vAa$ is reachable by a run that is a banal sequence. Another
pleasant property is that banal sequences can be explored in
logarithmic space: we need only to remember the current state of the
unique process that is active. We call \emph{configuration} a pair
$(s,\vect)$ consisting of a state $s$ of $\Bb$ and a global state $\vect$ of
$\Aa$. For convenience, we say that a configuration $(s,\vect)$ is
reachable by some sequence $\rho$ of transitions of $\Aa$ if $\rho$ leads
to $\vect$ from the initial state of $\Aa$, and if $s$ is reached 
in $\Bb$ from the initial state by the sequence of actions associated
with $\rho$ (this is well-defined since $\Bb$ is deterministic).  Note
also that we can explore any configuration
$(s,\vect)$ that is reachable by some banal sequence in
logarithmic space.  Let us call such pairs \emph{banally-reachable
  configurations}.

The first necessary condition for $\Bb$ being bisimilar to
$\Aa_1\otimes\dots\otimes\Aa_n$ is that for every banally-reachable
configuration $(s,\vect)$ the same actions are possible from $s$ and
$\vect$. This can be checked in logarithmic space as it is  easy to
verify its negation within this bound.

The second necessary condition is that every reachable configuration
is banally-reachable. Indeed, if $(s,\vect)$ is reachable by a
sequence that is not banal then the banal sequence $\rho$ obtained by
ordering the transitions process-wise also reaches
$\vect$. If a bisimulation exists then we are
guaranteed that $\rho$ reaches $s$ in $\Bb$.  This is because the state
reached by $\rho$ must be bisimilar to $s$, and $\Bb$ is minimal with
respect to bisimulation.

To show that one can check in logarithmic space that every reachable
configuration is banally-reachable, we consider
the negation of this property. We can then use the fact that
\LOGSPACE\ is closed under complement. We want to find a reachable
configuration that is not banally-reachable. If one exists then we can
look at one that is reachable in a shortest number of steps. This
means that there must exist a banally-reachable configuration
$(s_1,\vect_1)$, an action $b$ and a process $i$ such that
$(s_2,\vect_2)$ is not banally-reachable, where $\d_\Bb(s_1,b)= s_2$
and $\vect_2$ is obtained from $\vect_1$ by taking transition $b$ of
process $i$. This can be checked as follows. One produces on-the-fly a
banal sequence, when the part of process $i$ is finished an extra
transition with letter $b$ is taken. This way we have two states, one
before taking $b$ and one after. We then continue constructing banal
sequences from the two states with transitions of processes $i+1$ up
to $n$. This way we have obtained two sequences which differ by the
action $b$ of process $i$, and we check that the two states reached by
$\Bb$ are different.

Together, the two conditions above are also sufficient for $\vAa$ and
$\Bb$ being bisimilar, hence the result.

\section{Conclusion}

We have shown an \EXPTIME\ lower bound for the composition of
services that are described as a fully asynchronous product of finite state
machines. Thus, we answer the question left open in \cite{BCGLM03}. 
Since our lower bound holds for the simplest parallel composition
operation one can think of (no synchronization at all), it also
applies to richer models, such as products with synchronization on
actions as in \cite{LarSch00} or communicating finite-state machines
(CFSM) as in \cite{hbcs03,fbs04}. It is easy to see that the
simulation of a finite-state machine by a CFSM $\Aa$ with bounded
message queues is in \EXPTIME, since the state space of $\Aa$ is
exponential in this case. Hence, this problem, as well as any of its
variants with some restricted form of communication, is
\EXPTIME-complete as well.

An interesting open question  is what happens if we allow in the
asynchronous product arbitrary many copies of each finite state
machine. That is, we suppose that an available service can be used
by an arbitrary number of peers. This question reduces to a bounded
variant  of the simulation of a finite state machine by a BPP, and its
decidability status is open.

\medskip

\emph{Acknowledgement:} We thank the anonymous referees  for interesting
comments and suggestions for improvement.

\end{document}